\documentclass{PoS}

\title{The non-perturbative BRST quartet mechanism\\ in Landau gauge QCD: \\
Ghost-gluon and ghost-quark bound states}

\ShortTitle{Non-perturbative BRST quartet mechanism: 
Ghost-gluon and ghost-quark bound states}

\author{\speaker{Natalia Alkofer}%
\\
Institut f\"ur Physik,
Karl-Franzens-Universit\"at,
Universit\"atsplatz 5,
A-8010 Graz, Austria\\
       E-mail: \email{natalia.alkofer@edu.uni-graz.at}}

\author{Reinhard Alkofer\\
Institut f\"ur Physik,
Karl-Franzens-Universit\"at,
Universit\"atsplatz 5,
A-8010 Graz, Austria\\
E-mail: \email{reinhard.alkofer@uni-graz.at}}

\abstract{The non-perturbative BRST quartet mechanism in infrared Landau gauge
QCD is presented. It is demonstrated that positivity violation for transverse
gluons allows to identify the gluon's non-perturbative BRST quartet. To describe
the respective BRST-daughter  state a truncated Bethe-Salpeter
equation  for the gluon-ghost bound state is investigated. An analogous
construction for quarks yields a truncated Bethe-Salpeter
equation  for the quark-ghost bound state.
The gluon-ghost bound state equation in two space-time 
dimensions has been numerically solved.}

\FullConference{Xth Quark Confinement and the Hadron Spectrum\\
		 8-12 October 2012\\
		 TUM Campus Garching, Munich, Germany}

\begin{document}

The  Landau gauge gluon propagator  violates positivity,  see {\it e.g.\/} Refs.\
\cite{Bowman:2007du}  and references therein. Due to this property  the
one-(transverse)gluon-state possesses a negative norm and is thus removed from
the $S$-matrix. This construction can be formalised by considering the BRST
cohomology but this does not tell us how the removal of BRST non-invariant
states occurs dynamically. Within the context of Landau gauge QCD we formulated
some first steps into this direction \cite{Alkofer:2011pe}.   The
non-perturbative character of the BRST quartets of the transverse gluon (quark)
is evident from the fact that they contain besides the transverse gluon (quark) only
bound states. Therefore we need a method to solve bound state equations in an
approximately symmetry-preserving way. 

The quantum state of one transverse gluon constitutes  a parent state in its
BRST quartet, a fact which allows to identify the other members. The role of
these bound states in covariantly gauge-fixed  Yang-Mills (YM) theory as well as
for the kinematical aspects of gluon confinement can then be elucidated based 
on purely algebraic considerations. For quarks we adopt a similar approach as
for the gluons in order to clarify by constructing a (hypothetical) quark BRST
quartet whether quarks are also positivity violating.


The BRST transformation $\delta_B$ is defined as:
\begin{equation}
\delta_B A^a_\mu \, =\,  \widetilde Z_3 D^{ab}_\mu c^b \, \lambda  \; ,  
\; \delta_B q\, = \,-  i g t^a \widetilde Z_1 \, c^a \, q \, \lambda \; , 
\; \delta_B c^a \, = \, - \, \frac{g}{2} f^{abc} \widetilde Z_1 
 \, c^b c^c \, \, \lambda \; ,
\;  \delta_B \bar{c}^a \, = \, B^a
\, \lambda \; , \;
\delta_B B^a \, = \, 0,  
\label{BRST}
\end{equation}
where $D^{ab}_\mu$ is the covariant derivative. The  parameter
$\lambda$ has ghost number $N_{\mbox{\tiny FP}} = -1$, and $B^a$ is the 
Nakanishi-Lautrup field. $\widetilde Z_1$ and $\widetilde Z_3$ are the
ghost-gluon-vertex and the ghost wave function renormalisation constants.
Following the usual notation  the starting (negative norm) state 
is called 1st parent, its BRST transform 1st daughter. The Faddeev-Popov 
conjugated state
of the latter is named 2nd parent, the corresponding BRST transform 2nd
daughter.
Eq.\ (\ref{BRST}) allows to conclude that in the respective BRST quartet 
the daughter state of the transverse gluon is a gluon-ghost bound state 
\cite{Alkofer:2011pe}. The second parent is  a gluon-antighost bound
state with its BRST transform being the second daughter state. This algebraic
structure guarantees the vanishing of $S$-matrix elements
between transverse gluons and BRST singlets, the latter being the physical 
colour-neutral states.

There is no definite result on positivity for quarks. Within  functional 
methods this is due to the sensitivity of the analytic structure of the quark
propagator to details of the quark-gluon vertex \cite{Alkofer:2003jj}. This
vertex is, on the other hand, also very strongly influenced by dynamical and/or
explicit chiral symmetry breaking \cite{Alkofer:2008tt}. Therefore 
the resulting dynamical quark mass 
generation is highly susceptible to details of the dynamics. Which
mechanism secures then that the BRST-quartet ghost-quark bound states are
degenerate with the quark states is completely unknown. This motivates to 
investigate the bound state equations for the  non-perturbative BRST
quartet states with the aim to gain insight into the underlying mechanisms.

\begin{figure}[th]
\centerline{\includegraphics[width=150mm]{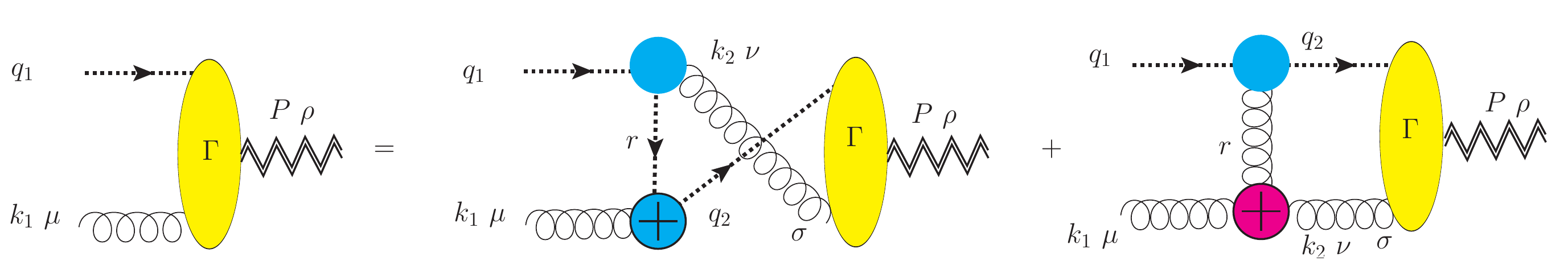}}
\caption{
Representation of the ghost-gluon Bethe-Salpeter equation.
Crosses denote dressed vertices.
\label{GhGl}}
\end{figure}

In Ref.\ \cite{Alkofer:2011pe} we derived the Bethe-Salpeter equations for  
ghost-gluon and  ghost-quark bound states in a truncation scheme based on an
infrared analysis. The gluon-ghost equation is
graphically represented in Fig.\ \ref{GhGl}, and in the following we will 
concentrate on this equation.
One notices that the gluon-ghost Bethe-Salpeter equation contains two terms,
a ghost exchange and a gluon exchange. Both are of the same infrared order 
because in the so-called scaling solution (for details of its properties
see Ref.\  \cite{Fischer:2008uz} and references therein) the fully dressed
three-gluon vertex is infrared divergent \cite{Alkofer:2004it}.
Employing the propagator
parametrisations of {\it e.g.\/} Ref.\ \cite{Alkofer:2003jj}, the ghost-gluon
vertex of Ref.\ \cite{Schleifenbaum:2004id}, and the three-gluon vertex of Ref.\
\cite{Alkofer:2008dt}, one arrives at a self-consistent equation for the
gluon-ghost Bethe-Salpeter amplitude with only known
quantities as input. 
As  demonstrated in Ref.~\cite{Alkofer:2011pe},
this bound state in the adjoint colour
representation is transverse, {\it i.e.}, it is as expected in the
massless representation of the Poincar{\'e} group. 
In the limit of soft momenta the ghost-gluon bound state is then described
by one amplitude $F(q^2)$ with $q$ being the relative momentum
between the gluon and the ghost.
A restriction to the presumably dominant ghost exchange and taking into account 
that the ghost-gluon vertex can be accurately approximated by the bare vertex
leads to a linear integral equation for $F(q^2)$ with only the 
gluon and ghost dressing functions, $Z(k^2)$ and $G(k^2)$, respectively, 
as input \cite{Alkofer:2011pe}. 

As recently propagator and vertex functions for two space-time dimensions 
became available \cite{Huber:2012zj} we solved the ghost-gluon equation 
in two dimensions,
\begin{eqnarray}
F(k_1^2) =  k_1^2  Z_3^2 g^2 N_c^2 
\int \frac {d^2k_2}{(2\pi)^2} \frac{G((k_1+k_2)^2)}{(k_1+k_2)^2}
\frac{G(k_2^2)}{k_2^2} \frac{Z(k_2^2)}{k_2^2}    
\left( 1 - \frac {(k_1\cdot k_2)^2}{k_1^2 k_2^2}\right) F(k_2^2) ,
\label{GhGlBSforF2}
\end{eqnarray}
and provided evidence for the existence of this bound state.
Details will be published elsewhere.

\paragraph*{Acknowledgments} 

~

We thank the organisers of the {\it Xth Quark Confinement and the Hadron
Spectrum} conference for all their efforts which made this extraordinary event
possible.
N.A.\ gratefully acknowledges financial support by the conference.
We thank M.\ Q.\ Huber for helpful discussions.

\end{document}